# Information Spread in a Connected World

ALON SELA, HILA OVED, IRAD BEN-GAL, Tel-Aviv University

## 1. INTRODUCTION

In the following work, we compare the spread of information by word-of-mouth (WOM) to the spread of information through search engines. We assume that the initial acknowledgement of new information derives from social interactions but that solid opinions are only formed after further evaluation through search engines. Search engines can be viewed as central hubs that connect information presented in relevant websites to searchers. Since they construct new connections between searchers and information in every query performed, the network structure is less relevant. Although models of viral spread of ideas have been inspected in many previous works [1], [2], [3], [4], [5], [6], [7], [8], [9], [10], [11], [12], [13], only few assume the acceptance of a novel concept to be solely based on the evaluation of the opinions of others [8], [5]. Following this approach, combined with that of models of information spread with threshold [1] that claim the propagation in a network to occur only if a threshold of neighbors hold an opinion, the proposed work adds a new theoretical perspective that is relevant to the daily use of search engines as a major information search tool. We continue by presenting some justifications based on experimentations. Last we discuss possible outcomes of over use of search engines vs. WOM, and suggest a hypothesis that such overuse might actually narrow the collective information set.

## 2. PROPOSED APPROACH

In the conventional model of viral spreading (known as Susceptible Infected Recover (SIR) model [10]), the probability of infection in a single encounter of an infected and a non-infected person is fixed. It is therefore referred as the "*Fixed*" model. The SIR model is modified to fit ideas spreading through word-of-mouth (WOM) by including group influence and is referred as the "*Group*" model. A second modification to fit ideas spreading through global search engines is referred as the "*Global*" model. Given these definitions, the formal notations of the models are presented below, followed by results of simulations. Last, a section presenting some empirical results is presented, with suggestions for practical use cases of these models.

### 2.1 Assumptions

Based on the SIR model excluding its recover state, the probability of infection changes from a fixed probability, to a varying probability according to the following assumptions:

1. **Herd Behavior:** Group dynamics influence the probability of acceptance of new ideas as people's attitudes tend to converge to the group norms [4]. The probability of adoption of a novel idea is linearly proportional to the number of people holding this idea in the relevant social group [8] [5].
2. **Global vs. Group Dynamics:** In the Group model, where information spreads primarily by word of mouth, one's social influential group consists of one's neighbors. In the Global model, where information spreads through the internet, it consists of the entire network.
3. **Unbiased search engines**: The search engines are unbiased; that is, that they present the user a sample of opinions that is proportional to the opinions that exist in the entire World Wide Web.

## 2.2 Formal Definition

Let $D = (V, A)$ be a directed graph with $n = |V|$ nodes and $m = |A|$ edges. Node $u$ is infected, e.g. accepting an idea at time $t$, if its state function $s_u(t) = 1$, otherwise $s_u(t) = 0$. An initial subset $V` \subseteq V$ is infected at $t=0$. The number of nodes in state $s_u(t) = 1$ is examined at a define time $t=T`$ as well as the time to spreads into k% of the network. Modifications of the SIR model were made to fit the Global and Group models mainly include social interactions. A Small World network [11] [12] [7] is constructed [13], [14] with a random subset $V` \subseteq V$ infected. When in the Fixed model, at every discrete time step, a random number from a uniform distribution $r_u^t(t) \sim U(0,1)$ is generated and compared to a constant threshold $\tau_c$. The comparison determines whether node $v$ passes the infection to node $u$ [1], [9]. The state function of $u$ in the Fixed model is thus $s_u(t+1) \equiv \begin{cases} 0 & r_u^t(t) > \tau_c \\ 1 & r_u^t(t) \leq \tau_c \end{cases}$. In the case of the Group social influence, the group of neighbors responsible for the social influence of $u$ is defined as the set of nodes having incoming edges to $u$, that is $\Gamma_u^+ \equiv \{u \in V : (v, u) \in A\}$. As in the case of the Fixed model, a random number is generated $r_u^t(t)$, and an infection occurs if the Group threshold $\tau_u^{gr}(t) \equiv \frac{\sum_{v \in \Gamma^+(u)} s_v(t-1)}{|\Gamma_u^+|} > r_u^t(t)$. In the Global model, the same equations hold, but the social influence group is redefined through the entire network set $A$. The Global threshold is thus the average infection rate over the entire network $\tau^{gl}(t) \equiv \frac{\sum_{v \in A} s_u(t-1)}{|A|}$ as illustrated in Figure 1 below.

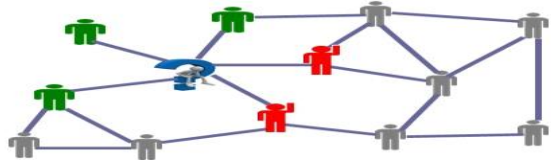

Figure 1: Computing the state of the person inside the question mark is as follows: This person has five neighbors; two of them are infected (in red and with a raised arm). In the Group model, the probability of him/her to become infected is p=2/5. In the Global model p= 2/13 (as the entire social group is of 13 nodes).

## 2.3 Results from Simulations

Simulations comparing the Global and the Group models [13], [14] reveals that in the Group models, the time until the epidemic starts is rather short, while in the Global model, the starting time is relatively long and highly unexpected. Figure 2 presents the average spreading times until the first 1% of the network is infected (Figure 2a) and the spreading times from 1% to 99% infection (Figure 2b). It can be seen that in the Group model, the starting of the spread is rather fast, while the Global model's start is significantly slower. The time for infection of 1%-99% of the network is rather similar.

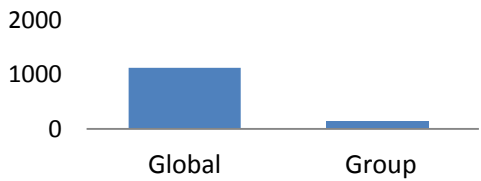

Figure 2a: Spreading time into 1% of the network for the two models: In the Group model, the spreading time is of ~120 time steps vs. the Global model, ~1100 time steps.

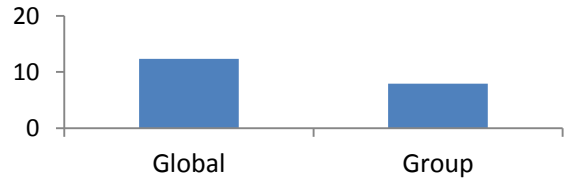

Figure 2b: Spreading time of two models: spread of 1%-99% of the network

## 2.4 Empirical Results to Validate the Assumptions

In order to validate the model`s assumptions with real data, a set of concepts that were queried using Google search engine. The collection of terms was made by using Google Correlate, a free tool that automatically extracts terms according to different time series that are given as an input. The set consisted of 60 terms that experienced a growth in the volume of searches over time that fits one of the three models mentioned above. The terms were randomly assigned to 5 different questionnaires, with 15 terms in each questionnaire and 5 terms from each model. Forty volunteers, 19 females and 21 males, answered the questionnaires, with a total of 544 non-empty responses. The respondents were asked to estimate the degree by which he/she would spread the novel term to a friend. Results of this experiment showed (Figure 3) that the model type (Group / Global / Fixed) affects the probability of spread (p-value = 0.02), as well as the gender (p-value = 2.4E-6), and the orientation to computers (p-value = 1.48E-8). Further inspection of the influence of the model type has revealed that the Group model and the Fixed model to be indistinguishable, thus the "model type" factor to result only from the difference between the Global model. Regarding the gender, females had a substantially higher probability of forwarding a message vs. males. Results indicate that terms emerging from the Global time series are more likely to be spread. The main reasons users claim they would forward the new terms were "*The term is novel and interesting*" (38%), "*The term is beneficial to the receiver*" (28%), "*The term is exciting.*" (11%), "*The term is hot gossip.*" (12%), "*The term is funny.*" (8%), and "*The term is connected to justice.*" (3%).

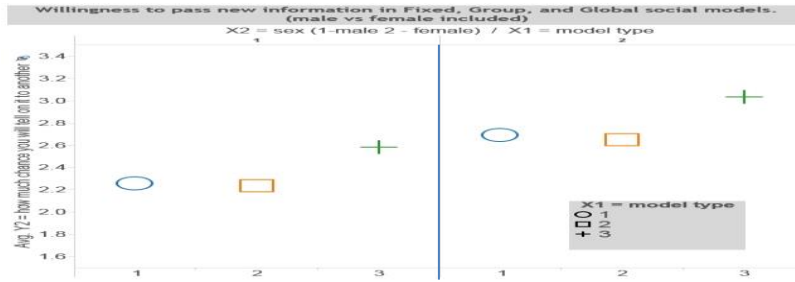

Figure 3: Willingness to spread a novel term deriving from the Fix model (1), Group model (2), Global model (3), by gender (right - female, left - male).

## 3. DISCUSSION

The terms that were found to be more viral emerged from the Global model. This model is based on an assumption that people accept novel ideas solely because their friends do so. It represents herd behavior on a global scale, and the simulations of its modeling predict slower spread of novel information. From our daily experience we all know that search engines dramatically increase the availability of information and assume them to be a tremendous source for information spreading. This contradiction might result from our unawareness of the lost information that has disappeared from our view. Search engines fetch surprisingly good answers for our queries. Nevertheless, we are never aware of the actual full set of web pages that globally fit our query. Furthermore, since most people do not tend to go beyond the first page of the search queries [21], the total number of viewed search results is even lower compared to WOM. From a viewpoint of "searcher", search engines dramatically increase information availability, but from the perspective of a person interested in spreading his/her novel ideas, our model proposes the advantage of publishing his/her novel idea through WOM methods including social media.